# Cortical Microcircuits from a Generative Vision Model

Dileep George*, Alexander Lavin, J. Swaroop Guntupalli, David Mely, Nick Hay, Miguel Lázaro-Gredilla

{dileep, alex, swaroop, david, nick, miguel}@vicarious.com

Vicarious AI, San Francisco, USA

**Abstract:**

**Understanding the information processing roles of cortical circuits is an outstanding problem in neuroscience and artificial intelligence. The theoretical setting of Bayesian inference has been suggested as a framework for understanding cortical computation. Based on a recently published generative model for visual inference (George et al., 2017), we derive a family of anatomically instantiated and functional cortical circuit models. In contrast to simplistic models of Bayesian inference, the underlying generative model's representational choices are validated with real-world tasks that required efficient inference and strong generalization. The cortical circuit model is derived by systematically comparing the computational requirements of this model with known anatomical constraints. The derived model suggests precise functional roles for the feedforward, feedback and lateral connections observed in different laminae and columns, and assigns a computational role for the path through the thalamus.**

## Introduction

Understanding information processing in visual cortical microcircuits is an unsolved problem in neuroscience. One avenue of research treats vision as generative model and derives cortical circuits from the inference mechanism in this generative model (Lee & Mumford 2003, George & Hawkins 2009). In our recent publication (George et al., 2017), we introduced the Recursive Cortical Network (RCN), a neuroscience-inspired probabilistic graphical model for vision that achieved state of the art results on several vision benchmarks with greater data-efficiency compared to prevalent deep neural networks. In this summary paper[1], we propose a biological implementation of RCN by combining its computational requirements with known anatomical and physiological constraints. We call this neural-RCN.

[1] This paper is accepted to CCN 2018. In a companion paper (Lavin et al., 2018), we use RCN to explain several visual phenomena. *Corresponding author.

While our model is consistent with the overarching idea of Bayesian inference and free energy minimization (Friston, 2010), in contrast to prior works that relied on simplistic models (Bastos et al., 2012), the inference algorithms and representational choices of RCN are validated with real-world tasks (George et al., 2017).

High level Bayesian inference frameworks that do not confront the problem of tractability in realistic settings run the risk of being overly general (Jones & Love, 2011), whereas testing on real world settings enable the discovery of architectural and algorithmic details that matter.

We focus on three aspects of RCN that were crucial for its performance – lateral connections, contour-surface interactions, and 'explaining away' – and derive the corresponding cortical microcircuits. These match several known details, and predict functional roles for several others.

## Recursive Cortical Network (RCN)

RCN is a structured probabilistic graphical model (PGM) for vision consisting of a contour hierarchy of features that interacts with a surface appearance canvas (Fig 1A). The contour hierarchy is learned as alternating layers of feature detectors, pools and lateral connections (Fig 1B). In Fig 1B, each circular node is a binary random variable, the elongated ellipses are categorical random variables, and the rectangles are factors that encode compatibility. Pooling provides invariance to local deformations, similar to the pooling in convolutional neural nets. The lateral connections, grey square 'factor nodes' in Fig 1 B&C, between the pools are learned to enforce contour consistency between the choices in adjacent pools. Fig 1C shows the hierarchical decomposition of a rectangle in terms of simple line segments at the bottom to more complex corner features at intermediate levels. Fig 1D is the graph corresponding to the representation of a letter "A" from a trained RCN.

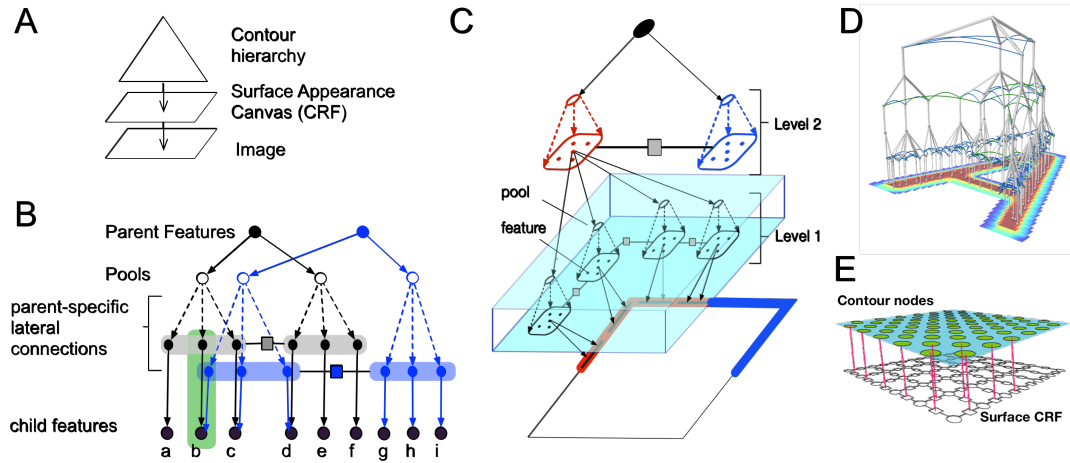

Figure 1: RCN probabilistic graphical model. See text for details.

The graphs corresponding to higher level features will share many of their lower level parts as shown in Fig 1B (blue and black), so that a hierarchy of objects is constructed out of many shared parts. The surface CRF, lower level of Fig 1E, encodes constraints regarding surface smoothness such that they are expected to vary smoothly when not interrupted by a contour (nodes in the upper layer), and discontinuously otherwise. A hierarchy consisting of convolutionally tiled graphs of different objects is a probabilistic model for the different objects in a scene.

Parsing is achieved by doing approximate MAP inference (inference to best explanation) using belief propagation (BP) (Pearl, 1988) with a schedule inspired by biology. A fast forward pass, which includes short-range lateral propagations, identifies nodes that are highly likely given the evidence. The backward pass focuses on highly active top-level nodes and includes longer range lateral propagations. The forward and backward passes assemble an approximate MAP solution that produces a complete segmentation of the input scene. See (George et al., 2017) for more details.

## Mapping of cortical circuits

Loopy BP computations are not directly represented in the PGM in Fig1, but can be understood as sending messages back and forth along the edges in the graph. Each edge has two messages going in opposite directions. Messages between binary variables are scalars representing the log likelihood ratio of the corresponding binary variable being ON, and messages between categorical variables are real-valued vectors.

In neural-RCN, inference computations corresponding to the features, pools and laterals of one level of the RCN maps to a specific cortical region. For example, Level 1 in Fig 1C would correspond to the primary visual cortex V1.

### Cortical column as a binary random variable

The core of neuro-RCN is viewing a cortical column as representing a single binary variable. In this interpretation, a cortical column a represents a 'feature' or a 'concept' -- for example, an oriented line segment in V1 or the letter 'B', in IT. The different laminae in a particular column correspond to the inference computations that determine the participation of this feature in different contexts: (1) laterally in the context of other features at the same level, (2) hierarchically in the context of parent features, (3) hierarchically as context for child features, and (4) pooling/un-pooling for invariant representations (Fig 2A)

**Neuronal clones:** RCN anticipates that a cortical column will contain "clones" of neurons that are nearly indistinguishable by their bottom-up input, but distinct when considering their lateral or top-down inputs. For example, RCN encodes higher-order lateral interactions in an efficient manner by having different copies of features for contours with different curvature Similar strategy is used for sequence representation and for border-ownership representation (George et al., 2017)

### Layer 2/3 lateral connections

Lateral factors in RCN encode an association field over contours such that lateral message propagation will tend to enhance smooth contours. In lateral propagation, the likelihood of each feature is calculated as a combination of bottom-up inputs from the features, and lateral messages from other pools. Layers 2 and 3 match the anatomical constraints for implementing these computations. They receive feed-forward inputs from the 'feature detector' layer-4 neurons (Harris & Shepherd, 2015), and send their axons across columns covering large distances and make patchy connections at their destinations (Binzegger et al., 2004).

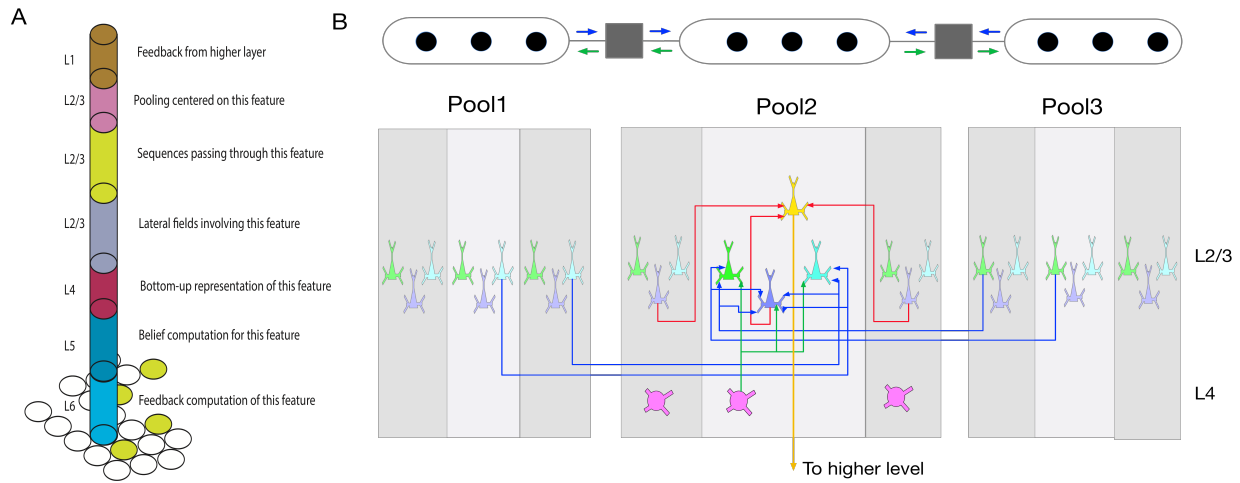

**Figure 2** A. Computational roles of different laminae in neural-RCN. B. Lateral connections in neural-RCN. See text for details.

A few aspects of the specific circuit (Fig 2B) predicted by RCN are noteworthy. Neural-RCN has separate neurons (green and cyan) within a column for receiving messages from a pool and for sending messages to that pool. A third neuron (purple) integrates the different inputs. The pooling neuron (yellow) pools the outputs from the purple neurons in multiple feature columns, and sends its output to the next level of the hierarchy. Having different neuron copies allows for segregation of incoming and outgoing messages, which is known to improve the accuracy of BP. However, a strict separation might not be required for reasonable performance.

The factor between the pools in RCN is a matrix that encodes the compatibility between the features in the different pools. In neural-RCN, this factor is implemented in the dendritic trees of the neurons involved. RCN stipulates the specific computations in the dendrites of the circuit in Fig 2B. For example, the green neuron that is receiving lateral axons from neighboring pool will first do a max-like operation over those activations and then add it (log domain) with the bottom up input it receives from layer 4 neurons.

Feedback computations are similar, and RCN predicts that a separate population of neurons in layer 2/3 or layer 5 performs this computation for the feedback pass. While the lateral connections are the same as in Figure 3, neurons in this population will have apical dendrites that extend to layer 1 to receive feedback from higher levels. Top-down messages act as a 'priors' on the pools at the lower level, and determine which pools in the children are ON/OFF. The specific feature column that is to be turned ON within a pool is then determined as the one most compatible with its neighboring pools, based on lateral message passing.

### Inter-blob and blob columns in V1

The use of a factorized contour-surface representation enables RCN to generalize to novel combinations of shapes and appearances. A similar segregation exists in V1 in terms of inter-blobs that represent oriented line segments, and the blobs that represent surface features like colors or textures (Sincich & Horton, 2005). RCN makes precise predictions about their interactions, based on the PGM in Fig 1E: The interactions between blobs (surface features) are gated by contour neurons in the inter-blob columns (potentially in layer 4). In Figure 3, the green-to-green lateral connections are the ones that represent surface continuity, and the red-to green lateral connections are the ones that represent a surface discontinuity. The specific prediction from RCN is that the contour neurons, using dendrite level inhibition and disinhibition (Stemmler et al., 1995), will select the appropriate lateral connections, as part of inference.

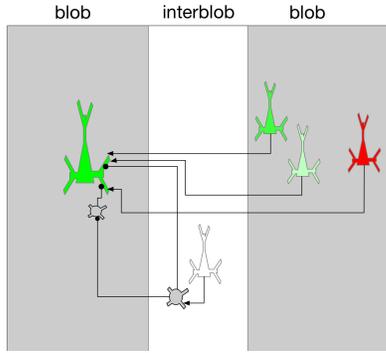

**Figure 3: Contour-surface interaction. See text for details.**

## Explaining away and top-down attention via the thalamus

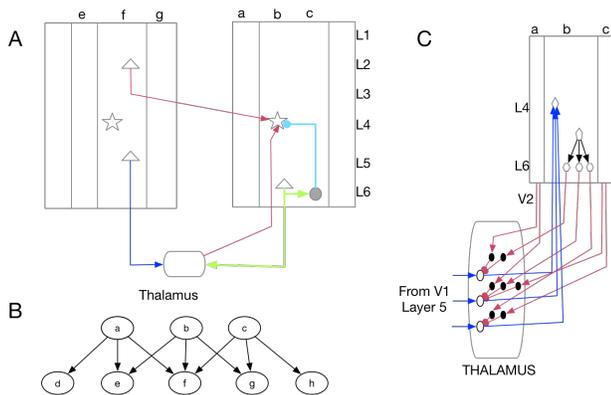

**Figure 4: Thalamus and explaining away. See text for details.**

An integrated functional role of the thalamic pathway (Rikhye et al., 2018) is an enduring mystery in neuroscience, and RCN makes predictions about this. Anatomical data show two feed-forward circuits: a direct cortico-cortical connection from from layer 2/3, and an indirect cortico-thalamo-cortical connection from layer 5. The thalamus also receives feedback connections from the higher level. The feedback projections from L6 also project back to L4 via an inhibitory circuit as shown in Fig 4A.

To understand RCN mapping, consider the PGM fragment in Figure 4B where the nodes a,b,c correspond to features at a higher level (V2) and nodes e,f,g, correspond to pools at a lower level (V1)(or it could represent the top-down connections from V1 to LGN.) Explaining away computations, in which the feed-forward messages from a child are affected by feed-back-messages that it has received, happen in child nodes that have more than one parent. This basic circuit can act as a template for understanding the pathway through the thalamus.

In neural-RCN, the direct cortical-cortical pathway provides fast feed-forward messages without explaining away. In the PGM of Figure 4B, The first feed-forward pass will assign equal strength the different competing hypotheses that have the same top-down prior. The pathway that goes through the thalamus includes explaining away and attention control. Maintaining these two pathways is advantageous because a fast feedforward pathway can alert the animal to novel situations that might be out of context. The inhibitory projection from L6 to L4 is an approximate version of this explaining away circuit as well, which provides faster but approximate explaining away mechanism. Figure 4C shows the detailed circuitry within the thalamus for explaining away computations.

## Discussion

Advances in neural imaging and recording technologies have led to a dense amount of data, but neuroscience as a field remains theory-sparse. How can we understand the cortex at a functional level? Our approach could offer a path forward. First build models whose representational choices are guided by neurobiology and real-world performance. Then work back from the model to make detailed connections to neurobiology. Through this cycle we hope to build better real-world models while simultaneously improving the precision and falsifiability of our neurobiological predictions.

## References

Bastos A.M et al.. (2012) Canonical microcircuits for predictive coding. Neuron.

Binzegger, T., Douglas, R. J., & Martin, K. A. (2004). A quantitative map of the circuit of cat primary visual cortex. Journal of Neuroscience, 24(39), 8441-8453.

Friston, K. (2010). The free-energy principle: a unified brain theory?. Nature Reviews Neuroscience, 11(2), 127.

George, D., Lehrach, W., Kansky, K., Lázaro-Gredilla, M., Laan, C., Marthi, B., ... & Lavin, A. (2017). A generative vision model that trains with high data efficiency and breaks text-based CAPTCHAs. Science, 358(6368)

George, D., Hawkins, J. (2009). Towards a mathematical theory of cortical microcircuits. PLoS Computational Biology.

Harris, K. D., & Shepherd, G. M. (2015). The neocortical circuit: themes and variations. Nature neuroscience, 18(2), 170.

Jones, M., & Love, B. C. (2011). Bayesian fundamentalism or enlightenment? On the explanatory status and theoretical contributions of Bayesian models of cognition. Behavioral and Brain Sciences, 34(4), 169-188.

Lee, T.S., Mumford, D., (2003). Hierarchical Bayesian Inference in the visual cortex. Journal of Optical Society of America.

Pearl, J. (1988). Probabilistic Reasoning in Intelligent Systems: Networks of Plausible Inference. Morgan Kaufmann.

Rikhye, R.V, Wimmer, R.D, Halassa, M.M. (2018). Toward an integrative theory of thalamic function. Annu. Rev. Neurosci. 41:163-83

Sincich, L. C., & Horton, J. C. (2005). The circuitry of V1 and V2: integration of color, form, and motion. Annu. Rev. Neurosci., 28, 303-326.

Stemmler, M., Usher, M., & Niebur, E. (1995). Lateral interactions in primary visual cortex: a model bridging physiology and psychophysics. Science, 269(5232), 1877-1880.

Zhou, H., Friedman, H. S., & Heydt, R. V. (2000). Coding of Border Ownership in Monkey Visual Cortex. J. Neurosci, 20(17), 6594-6611.